\documentclass[aps,prd,twocolumn,showpacs,preprintnumbers,nofootinbib]{revtex4}
\usepackage{amssymb,latexsym}
\usepackage{amsmath,amsbsy}
\usepackage{epsfig,bm}
\DeclareMathOperator{\pperp}{\mathbf{p}^{\perp}}
\DeclareMathOperator{\Pperp}{\mathbf{P}^{\perp}}

\DeclareMathOperator{\Pplus}{P^{+}}

\DeclareMathOperator{\kperp}{\mathbf{k}^{\perp}}
\DeclareMathOperator{\kplus}{k^{+}}

\DeclareMathOperator{\kminus}{k^{--}}

\DeclareMathOperator{\dperp}{\mathbf{\Delta}^\perp}
\DeclareMathOperator{\ie}{i \epsilon}
\DeclareMathOperator{\g5}{\gamma^5}
\DeclareMathOperator{\gp}{\gamma^+}
\DeclareMathOperator{\lamp}{\lambda^\prime}
\DeclareMathOperator{\dw}{D_{\text{W}}}
\DeclareMathOperator{\kpr}{\mathbf{k}^\perp_{\text{rel}}}
\DeclareMathOperator{\tr}{\text{Tr}}
\DeclareMathOperator{\res}{\text{Res}}
\DeclareMathOperator{\kpperp}{\mathbf{k}^{\prime \perp}}
\DeclareMathOperator{\xpp}{x^{\prime \prime}}
\DeclareMathOperator{\xp}{x^\prime}
\DeclareMathOperator{\Pbar}{\bar{P}}

\DeclareMathOperator{\xt}{\tilde{x}}

\begin{document}
\preprint{NT-UW 02-37}
\title{Generalized parton distributions and double distributions for $q\bar{q}$ pions}
\author{B.~C.~Tiburzi}
\author{G.~A.~Miller}
\affiliation{Department of Physics  
	University of Washington      
	Box 351560
	Seattle, WA 98195-1560}
\date{\today}

\begin{abstract}
We consider two simple covariant models for pions (one with scalar and the other with spin-$\frac{1}{2}$ constituents). 
Pion generalized parton distributions are derived by integration over the light-cone energy.
The model distributions are consistent with all known properties of generalized parton distributions, 
including positivity. We also construct the corresponding double distributions by appealing to Lorentz invariance. 
These ostensibly constructed double distributions lead to different generalized parton distributions that need not 
respect the positivity constraints. This inconsistency arises from the ambiguity inherent in defining double distributions
in the standard one-component formalism (even in the absence of the Polyakov-Weiss term). We demonstrate
that the correct model double distributions can be calculated from non-diagonal matrix elements of twist-two operators. 
\end{abstract}

\pacs{13.40.Gp, 13.60.Fz, 14.40.Aq}

\maketitle

\section{Introduction}

In recent years generalized parton distributions (GPDs) \cite{Muller:1998fv} have generated a considerable amount of attention. 
These distributions stem from hadronic matrix elements that are both non-diagonal with respect to hadron states and involve quark 
and gluon operators separated by a light-like distance. Thus physics of both inclusive (parton distributions, e.g.) and exclusive 
(form factors, e.g.) reactions is contained in the GPDs. At the leading-twist level, these new structure functions describe the soft 
physics of a variety of hard exclusive processes (see the reviews \cite{Ji:1998pc}). 

Since light-like correlation functions are involved in the description of deeply virtual Compton scattering (DVCS), e.g., there 
exists a simple decomposition of these matrix elements in terms of the light-cone Fock space wave functions of the initial and 
final states \cite{Brodsky:2001xy,Diehl:2000xz}. This representation of GPDs is ideal for physical intuition, however, comparatively little
has been done to show that the light-cone wave function representation is consistent with the reduction properties required of the 
generalized parton distributions. Below we undertake a simpler task of presenting covariant models for the pion which respect the 
properties of GPDs. Albeit simple, these models illustrate the utility and physicality of the light-cone Fock representation 
as well as provide a guide to understanding how the reduction relations arise in this formalism which will be useful when 
non-perturbative solutions for the Fock components in QCD become available. The scalar constituent model which we consider 
is merely the triangle graph 
with point-like vertices (this has been considered in \cite{Pobylitsa:2002vw}, see also \cite{Theussl:2002xp}). 
The spin-$\frac{1}{2}$ model is based on an earlier exactly soluble, $(1+1)$-dimensional light-front
 model \cite{Sawicki:hs}. Extension of this model to $(3+1)$ dimensions, which involves regularizing the divergent light-front current,
has been done in \cite{Bakker:2000pk}. 

Another approach is to use the formalism of double distributions (DDs) \cite{Radyushkin:1997ki,Radyushkin:1998es}.
This formalism elegantly explains the polynomiality conditions required of GPDs and thus gives one the ability to construct 
models consistent with known properties---although insight into model construction has often been limited to factorization \emph{Ans\"atze}.
Recently two-body light-front wave function models of the 
pion have been used to obtain GPDs \cite{Mukherjee:2002gb} based on DDs. Without modifying the quark distribution, 
the resulting GPDs violate the positivity constraints \cite{Tiburzi:2002kr}. This inconsistency is attributed to missing
contributions from non-wave function vertex diagrams which are absent when one uses non-covariant vertices. 
In general these diagrams are a substitute for higher Fock components, see \cite{Tiburzi:2002mn}. The covariant models used here, however, 
allow us to test the uniqueness of this ostensible construction.

Indeed we find that appealing to Lorentz invariance is only enough to determine one component of the
double distribution in the two-component formalism (even for $C$-odd distributions, where the Polyakov-Weiss $D$-term 
\cite{Polyakov:1999gs} is absent).
We show that in the scalar constituent model, missing the second component leads to incorrect GPDs. 
The same is true for the spin-$\frac{1}{2}$ constituent model, where additionally the positivity constraint is violated. 
The correct DDs can be calculated from non-diagonal matrix elements of twist-two
operators which we demonstrate for both models. On the other hand, exploiting the ambiguity inherent in 
defining DDs (which is akin to gauge freedom \cite{Teryaev:2001qm}) one can generate infinitely many different GPD models which share 
the same form factor \emph{and} quark distribution as well as satisfy polynomiality (and likely positivity).
  
The organization of the paper is as follows. First in section \ref{scalar}, we explicitly derive the GPD for the scalar triangle
diagram with point-like vertices. Next we show the DD for this model extracted from the form factor in the Drell-Yan frame 
leads to incorrect GPDs. Having encountered this problem, we review definitions of the double distributions in section \ref{twist}. 
Here we also calculate the correct DD for the scalar triangle diagram. In  \ref{spinor}, we present the model 
for the spin-$\frac{1}{2}$ case.
Next in section \ref{form}, we regularize the current and then extract this model's GPD.
Although not manifest, this model satisfies polynomiality, which is demonstrated in section \ref{prop}.
Using \cite{Mukherjee:2002gb} as a guide, we construct the DD for this model in section \ref{double}. 
Similar to section \ref{scalar}, this one-component DD too gives rise to a different GPD than the light-front projection.  
Additionally positivity is not satisfied by this one-component DD (section \ref{pc}). 
We calculate the missing component of the DD from matrix elements of twist-two operators in section \ref{real}. 
Lastly we conclude with a brief summary (section \ref{summy}).

\section{Pion with scalar constituents}\label{scalar}

For the pion model with scalar constituents, we choose the point-like Bethe-Salpeter vertex $\Gamma(k,P) = 1$, where
the coupling constant is assumed to be absorbed into the overall normalization. Furthermore, we choose derivative coupling
of the photon to charged scalar particles. 

\begin{figure}
\begin{center}
\epsfig{file=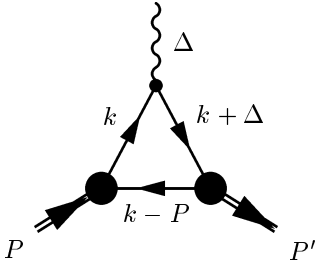}
\caption{Covariant triangle diagram for the pion electromagnetic form factor}
\label{ftri}
\end{center}
\end{figure}

The pion electromagnetic form factor for this model can be calculated from the Feynman triangle diagram. In order to derive
the GPD, however, we need to choose the kinematics specified in Figure \ref{ftri} with $k$ as the momentum of the struck 
quark. Using the stated pion vertex and taking the plus-component of the current\footnote{For any vector $a^\mu$, we define the light-cone variables $a^\pm \equiv \frac{1}{\sqrt{2}}( a^0 \pm a^3)$.}, we have
\begin{multline} \label{sff}
F(t) = \frac{-i |N|^2}{(1 - \zeta/2)P^+} \int d^4k (2 k^+ + \Delta^+) \big[ k^2 - m^2 + \ie\big]^{-1} \\
\times \big[ (k+\Delta)^2 - m^2 + \ie\big]^{-1} \big[ (P-k)^2 - m^2 + \ie\big]^{-1},
\end{multline}
where the momentum transfer is $t = \Delta^2$ and the skewness is defined relative to the initial state $\Delta^+ = - \zeta \Pplus < 0$.
Physically $\zeta$ plays the role of Bjorken variable for DVCS. Additionally we work in the frame where $\mathbf{P}^\perp = 0$. 

To turn Eq.~\eqref{sff} into an expression for the GPD $H(x,\zeta,t)$, 
we insert $\delta(k^+/ P^+ - x)$ to fix the momentum of the struck quark and keep $\zeta \neq 0$. This forces
\begin{equation} \label{defn}
F(t) = \int  H(x,\zeta,t) dx.
\end{equation}
Lastly we integrate over $\kminus$ to project onto the light cone. 
Doing the contour integration to extract $H(x,\zeta,t)$ in Eq.~\eqref{sff}, we are confronted with the poles
\begin{equation}
\begin{cases}
	k^-_{a} = k^-_{\text{on}}  -  \frac{\ie}{x}\\
	k^-_{b} = P^- + (k-P)^-_{\text{on}}  -  \frac{\ie}{x-1}\\	
	k^-_{c} = - \Delta^- + (k + \Delta)^-_{\text{on}}  -  \frac{\ie}{x^\prime}\\
\end{cases},
\end{equation}  
where the on-shell energies are $p^-_\text{on} = \frac{\pperp^2 + m^2}{2 p^+}$ and the abbreviation 
$x^\prime = \frac{x - \zeta}{1-\zeta}$ is used. Thus the non-vanishing contribution to the integral is
\begin{equation}
2 \pi i \; \theta[x(1-x)] \Big( \theta(x - \zeta) \res(k_b^-)  - \theta(\zeta - x) \res(k^-_a) \Big),
\end{equation}
which leads to
\begin{multline} \label{sgpd}
(1 - \zeta/2) H(x,\zeta,t) = \theta(x - \zeta) H_1(x,\zeta,t) \\ + \theta(\zeta - x) H_2(x,\zeta,t). 
\end{multline}
Using $\mathbf{k}^{\prime \perp} = \kperp + (1-x^\prime) \dperp$ for the relative transverse momentum of the final state, 
the functional forms are
\begin{align}
\begin{split}
H_1(x,\zeta,t)  & = (2 x - \zeta) |N|^2 \int d\kperp \dw(x,\kperp|M^2_\pi) \\  
& \quad \quad \times \dw(\xp,\kpperp|M^2_\pi)/ x(1-x) \xp 
\label{H1}
\end{split}\\
\begin{split} 
H_2(x,\zeta,t)  & = (2 x - \zeta) |N|^2 \int d\kperp \dw(x,\kperp|M^2_\pi) \\  
& \times \dw(\xpp,\mathbf{k}^{\prime\prime\perp}|t)/\zeta \xpp (1 - \xpp)(1-x) \label{H2},
\end{split}
\end{align}
where $\xpp \equiv x / \zeta$ and $\mathbf{k}^{\prime\prime\perp} \equiv  \kperp + \xpp \dperp$ are the relative momenta
of the photon. Additionally, we use the replacement 
\begin{equation}
\dw (x, \kperp | M^2)^{-1}  = M^2 - \frac{\kperp^2 + m^2}{x(1-x)},
\end{equation}
which is the propagator of the Weinberg equation \cite{Weinberg:1966jm}.

Comments about the GPD $H(x,\zeta,t)$ in Eq.~\eqref{sgpd} are in order. Firstly, the model is covariant and thus the 
sum rule and polynomiality conditions are met (see section \ref{prop} below for clarification). We have checked this
explicitly and suitable discussion can be found in \cite{Pobylitsa:2002vw,Theussl:2002xp}. Secondly, $H_1(x,\zeta,t)$
appearing in Eq.~\eqref{H1} satisfies the relevant positivity constraint for a compound scalar of 
scalar constituents (which appears in the Appendix of \cite{Tiburzi:2002kr}) which is clear from inspection.

Consideration of this model was first done from the perspective of DDs, see e.g.~the toy model of \cite{Radyushkin:1997ki}. 
This DD model was revisited recently with derivative coupling at the photon vertex in the Appendix of \cite{Mukherjee:2002gb}
and the same DD also appears in \cite{Pobylitsa:2002vw}. To derive the DD for this simple model, we appeal to Lorentz invariance, 
recalling along the way the relevant properties of DDs.  

First consider the form factor. In the $\zeta = 0$ frame, Eqs.~(\ref{H1}-\ref{H2}) reduce to the Drell-Yan formula \cite{Drell:1969km} 
via the definition in Eq.~\eqref{defn}
\begin{multline} \label{sFF}
F(t) =  |N|^2 \int \frac{dx d\kperp}{x(1-x)} \dw(x,\kperp|M^2_\pi) \\ \times  \dw(x,\kperp + (1-x) \dperp|M^2_\pi),
\end{multline}
where $t = - \dperp^2$. The form factor is Lorentz invariant and has a decomposition in terms of the Lorentz invariant
DD $F(x,y;t)$, namely
\begin{equation} \label{DDform}
F(t) = \int_0^1  \int_0^1  F(x,y;t) dy dx.
\end{equation}
The DD satisfies the following relations \cite{Radyushkin:1998es}: support property
\begin{equation} \label{A}
F(x,y;t) \propto \theta(1-x-y),
\end{equation}
reduction to the quark distribution at zero momentum transfer
\begin{equation} \label{B}
q(x) = \int_0^{1-x} F(x,y;0)  dy
\end{equation}
and is \emph{M\"unchen} symmetric \cite{Mankiewicz:1997uy}
\begin{equation}
F(x,y;t) = F(x,1-x-y;t).        \label{C}
\end{equation}
Using Eq.~\eqref{sFF}, we can write $F(t)$ in the form \eqref{DDform} with
\begin{equation} \label{sDD}
F(x,y;t) = \frac{x |N|^2 \theta( 1 - x - y)}{m^2 - x(1-x) M^2_\pi - y(1-x-y) t},
\end{equation}
which satisfies Eqs.~(\ref{A}-\ref{C}). The ingenuity of DDs comes about when we construct the GPD via
\begin{equation} \label{prescrip}
H(x,\zeta,t) = \int_0^1  \int_0^1  F(z,y;t) \delta(x - z - \zeta y ) dy dz.
\end{equation}
In this form the sum rule and polynomiality conditions follow trivially.
\begin{figure}
\begin{center}
\epsfig{file=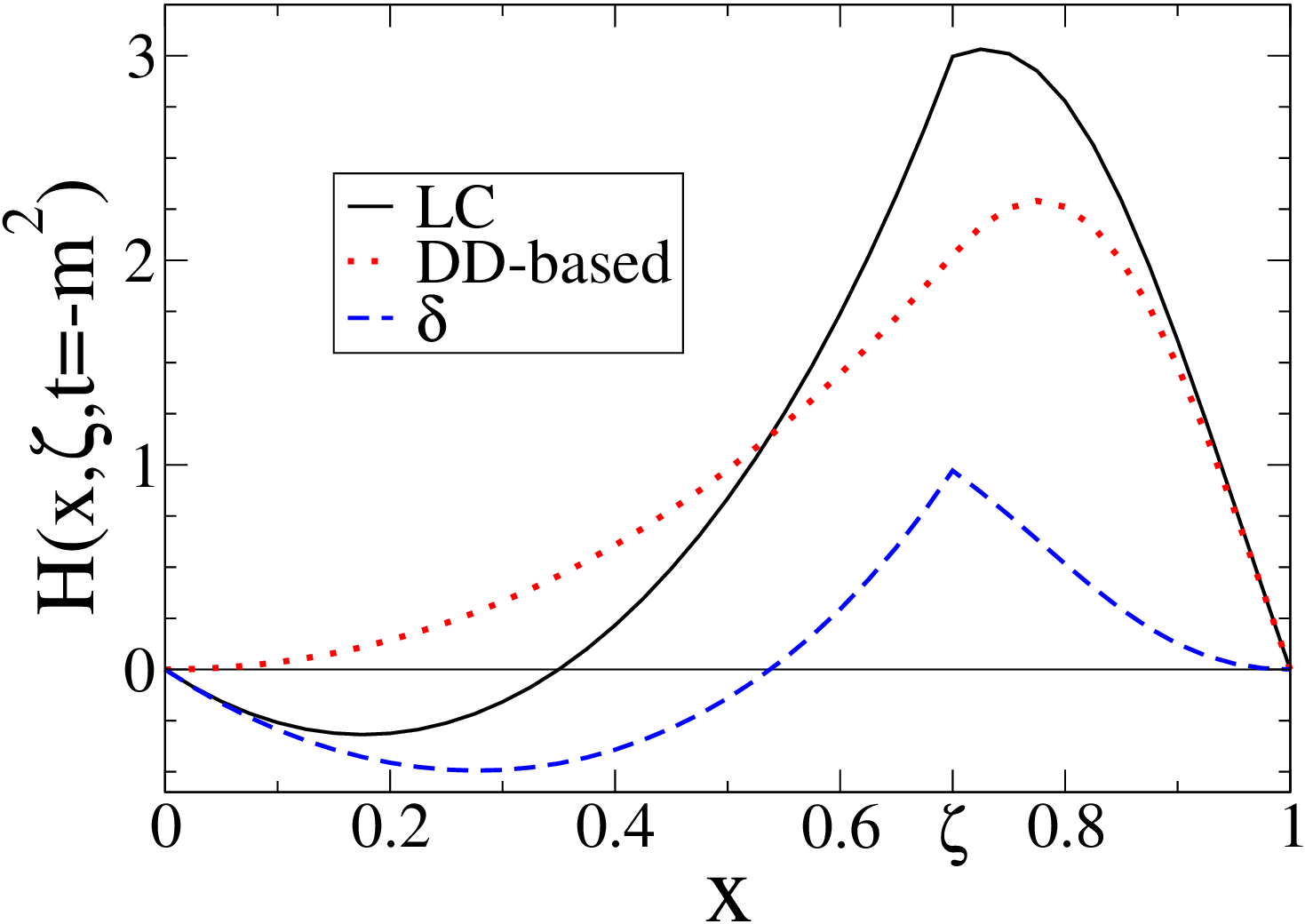,width=2.5in,height=1.5in}
\caption{Comparison of covariant GPDs for the scalar triangle diagram. The GPDs Eq.~\eqref{sgpd} (denoted LC) and 
Eq.~\eqref{prescrip} (DD-based) are plotted as a function of $x$ for fixed $\zeta = 0.7$ and $t = - m^2$ for the mass $M_\pi = m < 2m$.
We also plot the difference between the two curves ($\delta$). The area under the curves is identically $F(-m^2)$ for LC and DD-based GPDs, 
and hence zero for their difference $\delta$.}
\label{surprise}
\end{center}
\end{figure}

In Figure \ref{surprise}, we plot the GPD Eq.~\eqref{sgpd} as well as the GPD derived from DD via Eq.~\eqref{prescrip}.
Surprisingly the two are different despite the fact both models are covariant and posses the same form factor and quark distribution.
Additionally we plot their difference ($\delta$) as a function of $x$. 

\begin{widetext}
\section{Defining double distributions}\label{twist}
In this section we define the DDs via their moments following the two-component formalism of \cite{Polyakov:1999gs,Teryaev:2001qm}. 
Focusing on the ambiguities inherent by defining DDs in this fashion, we will understand why the DD in Eq.~\eqref{sDD} 
leads to an incorrect GPD. Moreover, we shall calculate the correct DD for the scalar triangle diagram from matrix elements of 
twist-two operators. We remark in passing that the two components of the DD (below $F$ and $G$, or $F$ and $D$-term in the standard
formalism) can be viewed as projections of a single function of two variables \cite{Belitsky:2000vk}. 

\subsection{Definitions}
The non-diagonal matrix elements of twist-two operators are more conveniently expressed in variables symmetric with respect to initial
and final states. To this end we define the average momentum $\Pbar^\mu = (P + P^\prime)^\mu / 2$. Let $\overset{\leftrightarrow}{D^\mu}
= \frac{1}{2} (\overset{\rightarrow}{\partial^\mu} - \overset{\leftarrow}{\partial^\mu} )$. Then for 
a pion of spin-$\frac{1}{2}$ constituents we have
\begin{multline} \label{decomp}
\langle \; \Pbar + \Delta/2 | \bar{\psi}(0) \gamma^{\big[\mu} i\overset{\leftrightarrow}{D}{}^{\mu_1}
\cdots i \overset{\leftrightarrow}{D}{}^{\mu_n}{}^{\big]} \psi(0) | \Pbar - \Delta/2 \; \rangle 
= \\ 2 \Pbar{}^{\big[ \mu} \sum_{k = 0}^n \frac{n!}{k!(n-k)!} A_{nk}(t) \; \Pbar{}^{\mu_1} \cdots \Pbar{}^{\mu_{n-k}} 
\Big(-\frac{\Delta}{2}\Big)^{\mu_{n - k + 1}} \cdots \Big(-\frac{\Delta}{2}\Big)^{\mu_n \big]}  \\
- \Delta^{\big[ \mu} \sum_{k = 0}^n \frac{n!}{k!(n-k)!} B_{nk}(t) \; \Pbar{}^{\mu_1} \cdots \Pbar{}^{\mu_{n-k}} 
\Big(-\frac{\Delta}{2}\Big)^{\mu_{n - k + 1}}  \cdots \Big(-\frac{\Delta}{2}\Big)^{\mu_n \big]},
\end{multline}
where the action of ${}^[ \cdots{}^]$ on Lorentz indices produces only the symmetric traceless part and $T$-invariance restricts 
$k$ in the first sum to be even and odd in the second. 
For a pion of scalar constituents, replace $\gamma^\mu$ with $2 i\overset{\leftrightarrow}{D^\mu}$. As it 
\end{widetext}
stands there is considerable freedom in the above decomposition,   e.g.~one could rewrite the above with 
\hbox{$k (n - k) B_{n,k-1}(t)$} as  a contribution to $A_{nk}(t)$. 
Carrying this out for all $k$, puts the bulk in the first term and 
renders the second term proportional only to the symmetric traceless part of ($n+1$) $\Delta$'s--- moments of the Polyakov-Weiss 
$D$-term \cite{Polyakov:1999gs}. This is the usually encountered form of the DD with $D$-term. Calculationally, however, we find
Eq.~\eqref{decomp} is the most useful.
 
Summing the moments, we produce the  DDs
\begin{align}
A_{nk}(t) & = \int_{-1}^{1} d\beta \int_{-1 + |\beta|}^{1 - |\beta|} d\alpha \; \beta^{n - k} \alpha^k F(\beta,\alpha;t) \label{Fdef}\\
B_{nk}(t) & = \int_{-1}^{1} d\beta \int_{-1 + |\beta|}^{1 - |\beta|} d\alpha \; \beta^{n-k}   \alpha^k G(\beta,\alpha;t) \label{Gdef}.
\end{align}
As a consequence of the restrictions on $k$ in the sums, the function $F(\beta,\alpha;t)$ is even in $\alpha$ 
while $G(\beta,\alpha;t)$ is odd. Also for $n$-even, there is no contribution from the $D$-term to the function $G(\beta,\alpha;t)$. 

These functions then appear in the decomposition of  matrix elements of light-like 
separated operators
\begin{multline} \label{DDdecomp}
\langle \; \Pbar + \Delta/2 \; | \; \bar{\psi}(-z^-/2) \; \rlap\slash z  \; \psi(z^-/2) \; | \; \Pbar - \Delta/2 \; \rangle 
\\ = 2 \Pbar \cdot z \int_{-1}^{1} d \beta \int_{-1 + |\beta|}^{1 - |\beta|} d \alpha \; 
e^{-i \beta \Pbar \cdot z + i \alpha \Delta \cdot z /2} F(\beta,\alpha;t) 
\\
- \Delta \cdot z \int_{-1}^{1} d\beta \int_{-1 + |\beta|}^{1 - |\beta|} d\alpha \; e^{-i \beta \Pbar \cdot z + i \alpha \Delta \cdot z /2}
G(\beta,\alpha;t),
\end{multline}
where $z^2 = 0$.

Denoting $\xi = - \Delta^+/ 2 \bar{P}^+$, the GPD in symmetric variables reads
\begin{multline}
H(\tilde{x},\xi,t) = \frac{1}{4\pi} \int dz^- e^{i \xt \Pbar^+ z^-} \\ 
\times \langle \; \Pbar + \Delta/2  |\bar{\psi}(-z^-/2) \gamma^+  
\psi(z^-/2) |  \Pbar - \Delta/2 \; \rangle. 
\end{multline}
Inserting Eq.~\eqref{DDdecomp} into this definition yields
\begin{multline} \label{JiGPD}
H(\xt,\xi,t) = \int_{-1}^{1} d\beta \int_{-1 + |\beta|}^{1 - |\beta|} d\alpha \; \delta(\xt - \beta - \xi \alpha) \\ 
\times
\Big[ F(\beta,\alpha;t) + \xi G(\beta, \alpha;t) \Big].  
\end{multline}
By integrating over $\xt$, we uncover \emph{two} sum rules: the sum rule for the form factor
\begin{equation} \label{Fsum}
\int d\beta \int d\alpha \; F(\beta,\alpha;t) = F(t)
\end{equation} 
and what we call the $G$-sum rule
\begin{equation} \label{Gsum}
\int d\beta \int d\alpha \; G(\beta,\alpha;t) = 0,
\end{equation}
which is trivial since $G$ is an odd function of $\alpha$. Eq.~\eqref{Gsum} has non-trivial consequences however, e.g.~it 
shows the method employed by \cite{Mukherjee:2002gb} leads only to the $F$ DD in Eq.~\eqref{DDdecomp}. This function 
integrates to the form factor via Eq.~\eqref{Fsum} and in the forward limit $\{\xi,t \to 0\}$ reduces to the quark distribution
(when integrated over $\alpha$). Thus $F(\beta,\alpha;t)$ should be properly termed the forward-visible DD, which encompasses 
\emph{more} than just neglecting the $D$-term. From Eq.~\eqref{decomp}, we see that $G(\beta,\alpha;t)$ does affect
higher moments of the GPD. This is the source of the discrepancy shown in Figure \ref{surprise} as we now demonstrate.

\subsection{Scalar model, revisited} 
To derive both $F$ and $G$ DDs for the scalar triangle diagram of section \ref{scalar}, we must now consider the action
of the operator $2 i \overset{\leftrightarrow}{D}{}^{\big[\mu} i\overset{\leftrightarrow}{D}{}^{\mu_1}
\cdots i \overset{\leftrightarrow}{D}{}^{\mu_n}{}^{\big]}$. This produces a factor
\begin{equation}
\frac{1}{2^n} (2 k + \Delta)^{\big[ \mu}(2 k + \Delta)^{\mu_1} \cdots (2 k + \Delta)^{\mu_n \big]}
\end{equation}
in the integrand of Eq.~\eqref{sFF}, which we now take in the symmetric frame. After the integration over $k$ is performed, 
we are left only with 
\begin{multline}
\int_{0}^{1} d\beta \int_{-1 + \beta}^{1 - \beta} d\alpha \;  D(\beta,\alpha;t)  \\ 
\times (2 \beta \Pbar - \alpha \Delta)^{\big[ \mu}(\beta \Pbar - \alpha \Delta/2)^{\mu_1}
\cdots(\beta \Pbar - \alpha \Delta/2)^{\mu_n \big]},
\end{multline}
where we have used the replacement
\begin{equation} \label{dba}
D(\beta,\alpha;t) = \frac{1}{m^2 - \beta (1-\beta) M^2_\pi - t [(1-\beta)^2 - \alpha^2]/4}.
\end{equation}

Using the binomial expansion, we can identify $F(\beta,\alpha;t)$ and $G(\beta,\alpha;t)$ via Eqs.~\eqref{Fdef} and \eqref{Gdef}, namely
\begin{align}
F(\beta,\alpha;t) & = \beta \; \theta(1 - \beta - \alpha) D(\beta,\alpha;t) |N|^2 \label{F1}\\
G(\beta,\alpha;t) & = \alpha \; \theta(1 - \beta - \alpha) D(\beta,\alpha;t) |N|^2  \label{G1}.
\end{align}
To compare with the results of section \ref{scalar}, we must revert to asymmetric variables which is accomplished
by $\{\beta \to x, y \to ( \alpha - \beta + 1)/2 \}$. The denominator common to both terms becomes
\begin{equation} \label{dxy}
D(x,y;t) = \frac{1}{m^2 - x(1-x) M^2_\pi - y (1-x-y) t}.
\end{equation}
Thus we have
\begin{align}
F(x,y;t) & = x \; \theta(1 - x - y) D(x,y;t) |N|^2 \label{F2}\\
G(x,y;t) & = (2 y + x - 1) \; \theta(1- x- y) D(x,y;t)  |N|^2 \label{G2}.
\end{align}
Notice the function $G(x,y;t)$ is \emph{M\"unchen} antisymmetric which is required because 
$G(\beta,\alpha;t)$ is odd with respect to $\alpha$.

To construct the GPD $H(x,\zeta,t)$ we must also convert Eq.~\eqref{JiGPD} to asymmetric variables.
\begin{multline}  \label{correct}
H(x,\zeta,t) = \int_0^1 dz \int_0^{1-z} dy \; \delta(x - z - \zeta y) \\
\times \Big[ F(x,y;t) + \frac{\zeta}{2 - \zeta} G(x,y;t) \Big]
\end{multline} 
We can now plot the GPD in Eq.~\eqref{correct} using the two DDs in Eqs.~\eqref{F2} and \eqref{G2}. 
The result agrees with Eq.~\eqref{sgpd} depicted in Figure \ref{surprise}. Moreover, the contribution 
from $\frac{\zeta}{2 - \zeta} G$ is identically the difference $\delta$ plotted in the figure.

\section{Pion with spin-$\frac{1}{2}$ constituents}\label{spinor}

\subsection{Bethe-Salpeter amplitude and pion wave function} \label{pion}
For the spin-$\frac{1}{2}$ model, we choose the trivially $q\bar{q}$ symmetric Bethe-Salpeter vertex
\begin{equation}
\Gamma(k,P) = - i g \g5.
\end{equation}
Here we have assumed only $\g5$ coupling at the quark-pion vertex with coupling constant $g$, 
whereas four Dirac structures exist \cite{Llewellyn-Smith:az}. 
This simple coupling is suggested by an effective interaction Lagrangian of the form 
(see, e.g.~\cite{Frederico:ye})
\begin{equation} \label{lint}
\mathcal{L}_{\text{I}} = - i g \; \bm{\pi} \cdot \bar{q} \g5 \bm{\tau} q, 
\end{equation}
where the coupling constant $g = m / f_{\pi}$, with $m$ the constituent mass and $f_{\pi}$ the pion decay constant. 
Notice the (ladder approximation) kernel is independent of light-cone time. Thus this model
(as well as the scalar triangle model in section \ref{scalar}) are special cases of the instantaneous formalism described by 
\cite{Tiburzi:2001je} in the impulse approximation. The relation of the vertex to the Bethe-Salpeter wave function is given by
\begin{equation} \label{bswfn}
\Psi(k,P) = \frac{i}{\rlap\slash k - m + \ie} \; (-ig) \gamma^5 \frac{i}{\rlap\slash k  - \rlap\slash P - m + \ie}.
\end{equation}

The valence wave function can be found by projecting the Bethe-Salpeter wave function onto the 
light-cone $x^+ = 0$, see e.g.~\cite{Liu:1992dg}. Using the normalization convention of \cite{Lepage:1980fj}, we have
\begin{multline} \label{val}
\psi(x,\kpr;\lambda,\lamp) = \frac{1}{2 P^+} \int \frac{d \kminus}{2 \pi} \frac{ \bar{u}_{\lambda} ( xP^+, \kperp ) }
{ \sqrt{x} } \gp \\ \times \Psi(k,P) \gp \frac{ v_{\lamp} \big( (1-x)P^+, \Pperp - \kperp \big) }{ \sqrt{1-x}},
\end{multline}
where $x$ is the fraction of the pion's plus momentum carried by the quark ($x = \kplus/\Pplus$), and the relative transverse momentum is $\kpr = \kperp - x \Pperp$. The valence wave function is found from Eq.~\eqref{val} to be
\begin{multline} \label{wfn}
\psi(x,\kperp;\lambda,\lamp) = \frac{g \sqrt{N_c/2} \; \mathcal{C}}{x(1-x)} \dw(x,\kperp|M^2_\pi) \\ \times
\Big[  k_{-\lambda} \; \delta_{\lambda,\lamp} - \lambda m \; \delta_{\lambda,-\lamp}  \Big] , 
\end{multline}
where we have employed the notation $k_{\lambda} = k^1 + i \lambda k^2$. 
As a result of the contour integration, we have a factor of $\theta[x(1-x)]$ implicitly in Eq.~\eqref{wfn}.
Additionally the wave function is symmetric under interchange of $x$ and $1-x$. As is known, the simplistic form of this wave function
leads to divergent quark distributions and form factors which will be handled below. Since this model is non-renormalizable, the choice of
regularization scheme influences the dynamics.

\subsection{Form factor and generalized parton distribution} \label{form}

The pion electromagnetic form factor for this model can be calculated from the Feynman triangle diagram. In order to extract the GPD, however, 
we need to choose the kinematics specified in Figure \ref{ftri} with $k$ as the momentum of the struck quark. As it stands, using the wave function
Eq.~\eqref{wfn}, the triangle diagram diverges.  Following the approach of \cite{Bakker:2000pk}, we covariantly smear the point-like 
photon vertex in Figure \ref{ftri} in a way reminiscent of Pauli-Villars regularization
\begin{equation} \label{ansatz}
\gamma^\mu \to \Gamma^\mu_\Lambda = \frac{\Lambda^2}{k^2 - \Lambda^2 + \ie} \gamma^\mu \frac{\Lambda^2}{(k+\Delta)^2 -\Lambda^2 + \ie}.
\end{equation}
Eq.~\eqref{ansatz} is a simple way to model non-$q\bar{q}$ components of the wave function. 
Alternatively one could smear the $q\bar{q}-\pi$ vertex in a covariant manner \cite{deMelo:1997cb,Jaus:zv}. This smearing should 
additionally respect the $q\bar{q}$ symmetry of the vertex. 
We do not pursue this option here since positivity constraints (see section \ref{pc}) are likely to be violated. 
On the other hand, one could use Pauli-Villars subtractions to regulate the theory, however, positivity would also be put into question. 
Because our concern is with model comparisons not phenomenology, we shall choose $\Lambda = m$ for simplicity. Although not obvious
from inspection, results for $\Lambda \neq m$ exhibit the same features investigated below. Most noteworthy, positivity remains 
satisfied when $\Lambda \neq m$. 

\begin{figure}
\begin{center}
\epsfig{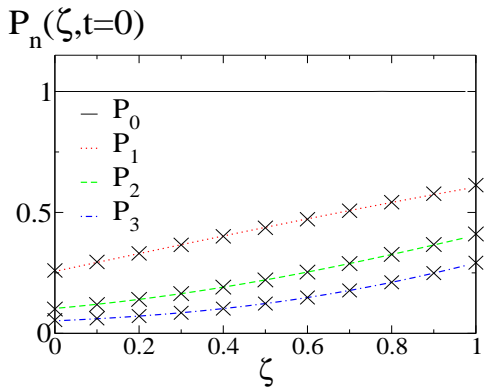}
\caption{Polynomiality conditions checked for the GPD Eq.~\eqref{gpd}. The moments $P_{n}(\zeta,t=0)$
from Eq.~\eqref{poly} are plotted as a function of $\zeta$ for $n=0,1,2,3$. Additionally X's denote the simple $(n+1)$-point
polynomial fit to the moment $P_n$.}
\end{center}
\label{figpoly}
\end{figure}

Considering matrix elements of the current operator $J^\mu$ between pion states, the model \eqref{ansatz} conserves current. 
This can be demonstrated most easily by calculating $\Delta \cdot J$ in the Breit frame. Additionally since the model is fully 
covariant, we can extract the electromagnetic form factor 
from any component of the current. In particular, potential zero-mode singularities present in matrix elements of $J^-$ have been removed 
by the photon vertex smearing  Eq.\eqref{ansatz} \cite{Bakker:2000pk}. Using the plus-component of the current, we have the expression
\begin{widetext}
\begin{equation} \label{ff}
F(t) =  \frac{i g^2 N_c |\mathcal{C}|^2 m^4}{1 - \frac{\zeta}{2}} \int \frac{d^4 k}{(2\pi)^4 P^+} \frac{\tr\Big[(\rlap\slash k + m )\g5  
(\rlap\slash k - \rlap\slash P  + m) 
\g5 (\rlap\slash k + \rlap\slash \Delta + m) \gp \Big]}
{\big[ k^2 - m^2 + \ie\big]^2 \big[ (k+\Delta)^2 - m^2 + \ie\big]^2 \big[ (P-k)^2 - m^2 + \ie\big]},
\end{equation}
where the momentum transfer is $t = \Delta^2$ and the skewness $\zeta$ is defined relative to the initial state: 
$\Delta^+ = - \zeta P^+ < 0$. 

To calculate the GPD, we follow the procedure described in section \ref{scalar}. The result can be written as
\begin{equation} \label{gpd}
(1 - \zeta/2) H(x,\zeta,t) = \theta(x - \zeta) H_{\text{eff}}(x,\zeta,t) 
+ \theta(\zeta - x) \Big( H_{\text{inst}}(x,\zeta,t) + H_{\text{nval}}(x,\zeta,t) \Big).
\end{equation}
where $H_{\text{eff}}$ is the piece determined by the effective two-body wave function, 
$H_{\text{inst}}$ is the contribution from instantaneous propagation of the spectator quark,
and the remaining contributions we term non-valence (although strictly speaking the instantaneous piece is 
also of the non-valence variety). It is a peculiarity of this model that explicit instantaneous terms
are not present for $x > \zeta$.

The functional forms are
\begin{align} \label{Feff}
 H_{\text{eff}}(x,\zeta,t)  & = \int \frac{d\kperp}{(2\pi)^3} \sum_{\lambda,\lamp} 
\psi^*_{\text{eff}}(x^\prime,\kpperp ;\lambda,\lamp) \psi_{\text{eff}}(x,\kperp;\lambda,\lamp),\\
 H_{\text{inst}}(x,\zeta,t) & =  - \mathcal{A} \int d\kperp \frac{4 \zeta \dw^3(\xpp,\kperp|t)}{(1-x)\xpp(1-\xpp)} \label{Finst}\\
 H_{\text{nval}}(x,\zeta,t) & =  - \mathcal{A} \int d\kperp \frac{2(\kperp \cdot \kpperp + m^2) \dw(x,\kperp|M_\pi^2) 
\dw^2(\xpp,\mathbf{k}^{\prime\prime\perp}|t)}{x(1-x)\xpp(1-\xpp)x^\prime(1-x^\prime)(1-\zeta)} \notag \\
& \times \Big[2\zeta \dw(\xpp,\mathbf{k}^{\prime\prime\perp}|t) + \dw(x,\kperp|M^2_\pi) \Big], 
\label{Fnval}
\end{align}
where we have defined the effective wave function
\begin{equation} \label{psieff}
\psi_{\text{eff}}(x,\kperp; \lambda,\lamp) = \frac{g \sqrt{N_c/2} \; \mathcal{C} m^2 }{x^2(1-x)}  
\Big[  k_{-\lambda} \; \delta_{\lambda,\lamp} - \lambda m \; \delta_{\lambda,-\lamp}  \Big] \dw^2(x,\kperp|M^2_\pi). 
\end{equation}
and made the abbreviation $\mathcal{A} = g^2 |\mathcal{C}|^2 N_c m^4 / 2 (2\pi)^3$.
It is sensible to think of Eq.~\eqref{psieff} as an effective wave function since $x \to 1-x$ symmetry has been lost. 
Moreover the wave function vanishes at $x = 1$ and is non-vanishing at $x = 0$. This is a desirable addition to the dynamics
stemming from the regularization. Notice the ladder kernel \eqref{lint} is momentum independent and hence does not vanish at $x = 0,1$. 
This is the dynamical reason why the un-regularized wave function Eq.~\eqref{wfn} does not vanish at the end points.  
Continuity of the GPD at $ x = \zeta$ follows directly from Eqs.~(\ref{Feff}-\ref{Fnval}). 

\begin{figure}
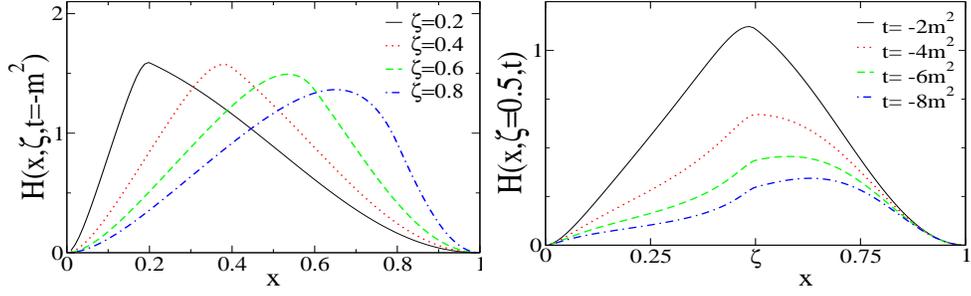

\begin{center}
\epsfig{file=lct.eps,height=1.5in,width=2.5in}
\epsfig{file=lcz.eps,height=1.5in,width=2.5in}
\caption{On the left, the GPD Eq.~\eqref{gpd} is plotted at fixed $t = - m^2$ for 
a few values of $\zeta$. On the right, the same GPD is plotted for fixed $\zeta = 0.5$ for a few values of $t$.
The model parameters are arbitrarily chosen as: $M = 0.14$ MeV and $m = 0.33$ MeV.}
\end{center}
\label{figLC}
\end{figure}

In Figure \ref{figLC}, we plot the GPD for the parameters: $M = 0.14$ MeV and $m = 0.33$ MeV. On the left, the graph
shows the GPD for a few values of $\zeta$ as a function of $x$ for fixed $t$, while on the right we have fixed $\zeta$ and $t$ varying. 

\subsection{Sum rule and polynomiality} \label{prop}
Since not manifest, one should check the covariance of the model Eq.~\eqref{gpd}. With the covariant starting point Eq.~\eqref{ff}, we 
anticipate polynomiality will be satisfied which provides a useful check on our expressions Eqs.~(\ref{Feff}-\ref{Fnval}). 
\end{widetext}
First we define the moments of the GPD with respect to asymmetric variables
\begin{equation} \label{poly}
P_n(\zeta,t) = \int x^n H(x,\zeta,t) dx.
\end{equation}
Polynomiality requires the moments $P_n$ to be of the form
\begin{equation} \label{polycon}
P_n(\zeta,t) = \sum_{j = 0}^{n} a_{n j}(t) \; \zeta^j.
\end{equation}
The zeroth moment is merely the sum rule for the form factor, hence $a_{0 0}(t) = F(t)$. 
For simplicity, we check a few of the lowest moments for the polynomiality condition
Eq.~\eqref{polycon} at $t = 0$. In Figure \ref{figpoly}, we plot the moments $P_n(\zeta,0)$
for $n = 0,1,2,3$ which appear as smooth functions. Additionally we plot simple $(n+1)$-point polynomial 
fits to the moments, which line up nicely with the integrals \eqref{poly}. 

\subsection{Double distribution} \label{double}
To construct the DD, we shall first proceed incorrectly by appealing to Lorentz invariance as in section \ref{scalar}. This will at least 
lead to one component of the DD, which can be compared to Eq.~\eqref{gpd}.

Using Eq.~\eqref{Feff} in the $\Delta^+ = 0$ (Drell-Yan) frame, we can write $F(t)$ in the form \eqref{DDform} with
\begin{multline} \label{DD}
F(x,y;t) =  \Big( 3 m^2 - M^2_\pi x(1-x) + y(1-x-y) t \Big) \\ \times 
\frac{2 \pi \mathcal{A} \; \theta(1-x-y)\;  y(1-x-y)}{(1-x) \Big[m^2 - M^2_\pi x(1-x) - y (1-x-y) t  \Big]^3}.
\end{multline}
Aside from factors arising from spin, this DD is basically the same as that considered \cite{Mukherjee:2002gb} 
which one can realize by utilizing $\lambda^2 = - M^2_\pi /4 + m^2$. Not surprisingly, then, 
this DD satisfies Eqs.~(\ref{A}-\ref{C}). For reference we give the quark distribution function 
\begin{equation} \label{q}
q(x) = \frac{2\pi \mathcal{A}}{6} \frac{(1-x)^2 \Big[ 3 m^2 - M^2_\pi x(1-x) \Big]}{[m^2 - M^2_\pi x(1-x)]^3},
\end{equation}
which could be calculated directly from $\psi_{\text{eff}}$ in Eq.~\eqref{psieff}.

\begin{figure}
\begin{center}
\epsfig{file=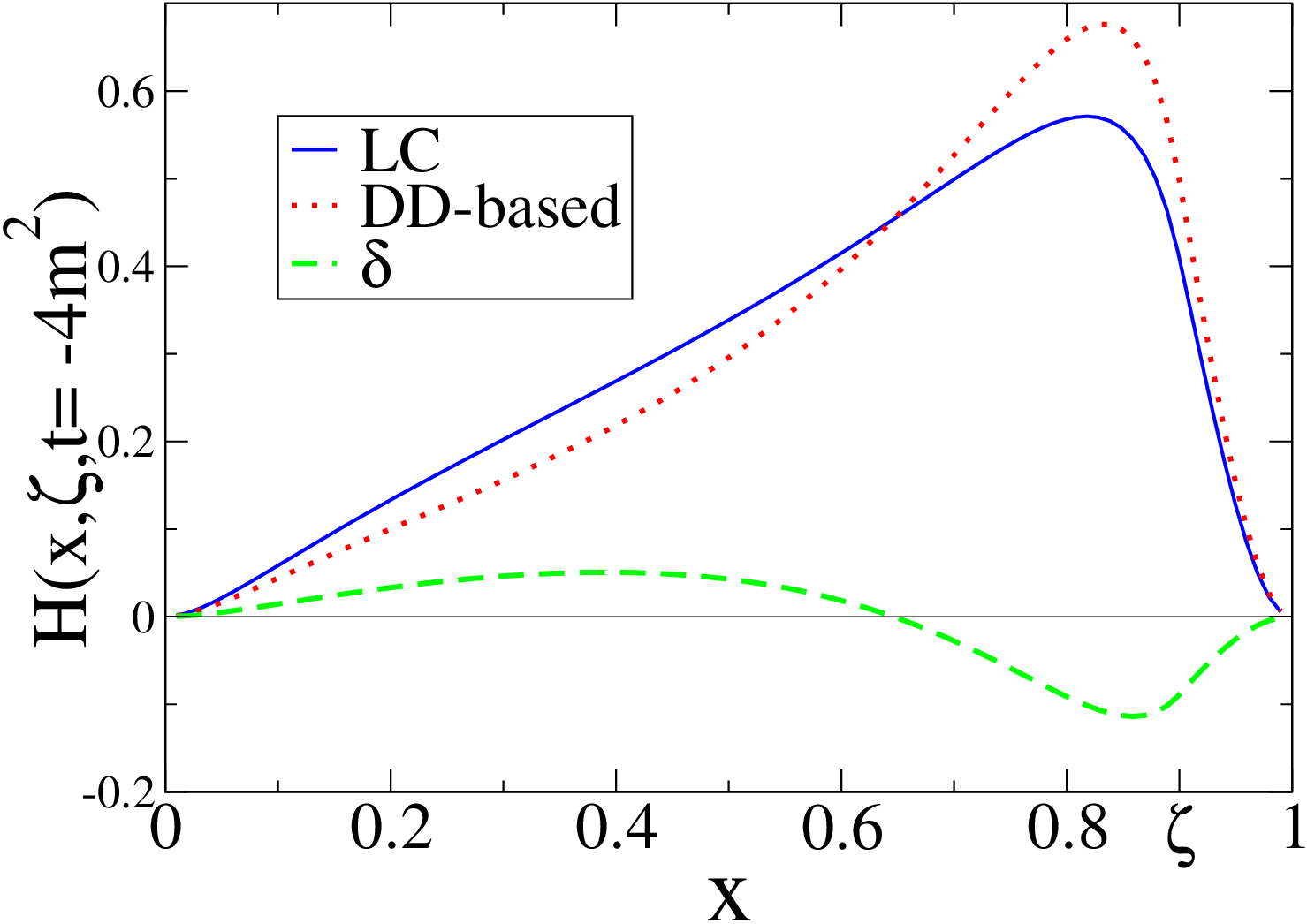,width=2.5in,height=1.5in}
\caption{Comparison of covariant GPDs for the spinor triangle diagram. The GPDs Eq.~\eqref{gpd} (denoted LC) and 
Eq.~\eqref{DD} (DD-based) are plotted as a function of $x$ for fixed $\zeta = 0.9$ and $t = - 4 m^2$ for the mass $M_\pi = 0.15 m$.
We also plot the difference between the two curves ($\delta$). The area under the curves is identically $F(-4m^2)$ for LC and DD-based 
GPDs, and hence zero for their difference $\delta$.}
\label{surprise2}
\end{center}
\end{figure}

In Figure \ref{surprise2}, we plot the GPD Eq.~\eqref{gpd} as well as the GPD derived from DD Eq.~\eqref{DD} via Eq.~\eqref{prescrip}.
As in section \ref{scalar}, the two are different despite the fact both models are covariant and posses the same form factor 
and quark distribution. Additionally we plot their difference as a function of $x$.

\subsection{Positivity Constraints} \label{pc}
Here we demonstrate another difference between the GPD in \eqref{gpd} and the one stemming from the  
the one-component DD Eq.~\eqref{DD}. To do so, we look at the positivity constraints. Originally these constraints appeared in 
\cite{Radyushkin:1998es,Pire:1998nw} and were derived from the positivity of the density matrix by restricting 
the final-state parton to have positive plus-momentum 
(and ignoring the contribution from $E(x,\zeta,t)$ for the spin-$\frac{1}{2}$ case).  Although the 
matrix elements involved for GPDs are off diagonal, they are still restricted by positivity and their diagonal elements. Correcting the constraints
for the presence of the $E$-distribution was first done in \cite{Diehl:2000xz}. By considering the positivity of the norm on Hilbert space, 
stricter constraints for the spin-$\frac{1}{2}$ distributions $H$ and $E$ have recently appeared as well as constraints for 
the full set of twist-two GPDs \cite{Pobylitsa:2001nt}. 

\begin{figure}
\begin{center}
\epsfig{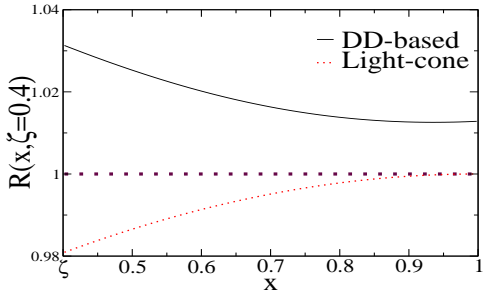}
\caption{Comparison of GPDs: GPD calculated from the DD Eq.~\eqref{DD} via Eq.~\eqref{prescrip} (DD-based) compared with
the light-cone projection of the form factor Eq.\eqref{gpd} (Light-cone) for fixed $\zeta =0.4$ at $t = 0$. Here we plot
the ratio $R(x,\zeta)$ appearing in Eq.~\eqref{posi} as a function of $x>\zeta$. Positivity constrains this ratio 
to be less than one.}
\end{center}
\label{figpos}
\end{figure}

For the scalar pion case there is of course no contribution from the non-existent $E$-distribution and hence the original bounds are actually correct 
(modulo factors due to the difference of a spin-$\frac{1}{2}$ proton versus a spin-$0$ pion). Given the matrix element definition 
of the GPD consistent with equation \eqref{defn}, namely
\begin{multline}
H(x,\zeta,t) = \frac{1}{1-\zeta/2}  \int \frac{dz^-}{4\pi} e^{i x P^+ z^-} \\ \times 
\langle \pi(P^\prime) | \bar{\psi}(0) \gp \psi(z^-) | \pi(P) \rangle,
\end{multline} 
the spin-$0$ positivity constraint (for $x>\zeta$) reads
\begin{equation} \label{posi}
R(x,\zeta) \equiv \big( 1 - \zeta/2 \big) \frac{\big|H(x,\zeta,0)\big|}{\sqrt{q (x) q(x^\prime)}}   \leq 1,
\end{equation}
where $q(x)$ is the model distribution function in Eq.~\eqref{q}. Of course the above result holds for finite $-t$, 
however since the function $\mathcal{F}$ decreases with $-t$, Eq.~\eqref{posi} is the tightest constraint. Notice
for $M^2_\pi \neq 0$, the limit $t = 0$ is in an unphysical region. If we treat this limit as formal, however, and analytically
continue our expressions, we can use Eq.~\eqref{posi}. Such continuation is consistent with the light-cone Fock space
representation of GPDs \cite{Brodsky:2001xy,Diehl:2000xz}.

Given the constraint Eq.~\eqref{posi}, we can test whether GPDs calculated from the light-cone projection
\eqref{Feff} and DD \eqref{DD} satisfy positivity. In Figure $6$, we plot $R(x,\zeta)$ for each 
GPD as a function of $x$ for the fixed value of $\zeta = 0.4$. There is noticeably different behavior in the figure: 
positivity is violated by the DD-based model. As above (section \ref{twist}), we must carefully derive contributions 
from the other component $G(x,y;t)$. 

\subsection{Derivation of the correct DDs}\label{real}
To derive both $F$ and $G$ DDs for the spin-$\frac{1}{2}$ pion model, we must consider the action of the
operator $\gamma^{\big[\mu} i\overset{\leftrightarrow}{D}{}^{\mu_1}
\cdots i \overset{\leftrightarrow}{D}{}^{\mu_n}{}^{\big]}$ between non-diagonal pion states. 
Inserted into Eq.~\eqref{ff} which is now taken in the symmetrical frame,  this operator produces
\begin{widetext}
\begin{equation} \label{nspinor}
\frac{i \mathcal{A}}{\pi} \int d^4 k \frac{\tr\Big\{(\rlap\slash k + m )\g5  
(\rlap\slash k - \rlap\slash \Pbar  + \rlap\slash \Delta / 2 + m) \g5 (\rlap\slash k + \rlap\slash \Delta + m) \gamma^{\big[ \mu} \Big\}
(k + \Delta/2)^{\mu_1} \cdots (k + \Delta/2)^{\mu_n \big]}}
{\big[ k^2 - m^2 + \ie\big]^2 \big[ (k+\Delta)^2 - m^2 + \ie\big]^2 \big[ (k - \Pbar + \Delta/2)^2 - m^2 + \ie\big]}.
\end{equation}
The presence of the trace
\begin{equation}
\tr\{ \ldots \}{}^\mu = 4 \Big[ \bar{P}{}^\mu (m^2 - k^2 - k \cdot \Delta) + \frac{\Delta^\mu}{2} (m^2 - k^2 + 2 k \cdot \Pbar) 
+ k^\mu (m^2 - k^2 - \frac{t}{2} + 2 k \cdot \Pbar - k \cdot \Delta)  \Big]
\end{equation}
complicates evaluating Eq.~\eqref{nspinor} by requiring contributions from diagrams reduced by one propagator. Since we
have smeared the photon via \eqref{ansatz}, the reduced diagrams are finite. Let us denote the propagators
simply by $\mathfrak{A} = (k - \Pbar + \Delta/2)^2 - m^2 + \ie$, $\mathfrak{B} = (k+\Delta)^2 - m^2 + \ie$ and  
$\mathfrak{C} = k^2 - m^2 + \ie$. To correctly evaluate Eq.~\eqref{nspinor}, we must write the trace as
\begin{equation} \label{redtrace}
\tr\{ \ldots \}{}^\mu = 4 \Big[ \bar{P}{}^\mu \big(\frac{t}{2} - \frac{1}{2} (\mathfrak{B} + \mathfrak{C}) \big) + 
\frac{\Delta^\mu}{2} \big( M^2 - \frac{t}{4} - \mathfrak{A} + \frac{1}{2} (\mathfrak{B} - \mathfrak{C}) \big) 
+ k^\mu ( M^2 - \frac{t}{2} - \mathfrak{A})  \Big]
\end{equation}
and evaluate each term separately canceling propagators in the denominator of Eq.~\eqref{nspinor}. 
These integrals can easily be evaluated using Feynman parameters. For example, let us consider the non-reduced 
contribution from Eq.~\eqref{redtrace}. The denominator of Eq.~\ref{nspinor} appears as
$\mathfrak{A} \; \mathfrak{B}^2 \; \mathfrak{C}^2$ and so we introduce two Feynman parameters $\{x,y\}$ 
to render the denominator specifically in the form $[x \mathfrak{A} + y \mathfrak{B} + (1-x-y) \mathfrak{C}]^{-5}$.
One then translates $k^\mu$ to render the integral (hyper-) spherically symmetric via the definition 
$k^\mu = l^\mu + \beta \Pbar{}^\mu - (\alpha + 1) \Delta^\mu / 2$. Here $\beta = x$ and $\alpha = x + 2y - 1$. Using a Wick rotation
to evaluate the resulting integral over $l$, we can cast the contribution to Eq.~\eqref{nspinor} from non-reduced terms
in the form
\begin{multline} \label{nonred}
\pi \mathcal{A} \int_{0}^{1} d\beta \int_{-1 + \beta}^{1 - \beta} d\alpha \frac{1}{4} 
[(1-\beta)^2 - \alpha^2] D(\beta, \alpha;t)^3
\Big( 2 \Pbar (\beta M_\pi^2 + (1-\beta)t/2) - \Delta \alpha (M_\pi^2 - t/2)  \Big)^{\big[ \mu}
\\ \times \sum_{k = 0}^n \frac{n!}{k! (n-k)!} \beta^{n-k} \alpha^k \Pbar{}^{\mu_1} \cdots \Pbar{}^{\mu_{n-k}} 
\Big(-\frac{\Delta}{2}\Big)^{\mu_{n - k + 1}} \cdots \Big(-\frac{\Delta}{2}\Big)^{\mu_n \big]}
\end{multline}
\end{widetext}
with $D(\beta,\alpha;t)$ given by Eq.~\eqref{dba}. Given the form of Eq.~\eqref{nonred}, we can identify $\{\beta,\alpha\}$
as DD variables and hence read off contributions to $F$ and $G$ DDs. 
\begin{align}
\begin{split}
\delta F(\beta,\alpha;t)  & = \pi \mathcal{A} \; [(1-\beta)^2 - \alpha^2] \; D(\beta,\alpha;t)^3  \\
&    \quad \quad \times \big(\beta M_\pi^2 + ( 1- \beta) t/2\big)
\end{split}\\ 
\begin{split}
\delta G(\beta,\alpha;t) & = \pi \mathcal{A} \; \alpha \; [(1-\beta)^2 - \alpha^2] \; D(\beta,\alpha;t)^3 \\
&  \quad \quad  \quad \quad \quad \times \big( M^2_\pi - t/2\big). 
\end{split}
\end{align}
Notice that these contributions respect the properties of DDs, namely $\delta F$ is even in $\alpha$ while
$\delta G$ is odd. This need not be the case, however, for each intermediate step of the calculation, e.g.~contributions
from $\mathfrak{B}$-reduced terms and $\mathfrak{C}$-reduced terms are individually neither even nor odd in $\alpha$ while 
their sum is even and difference is odd.

Ignoring for the moment contributions from $\mathfrak{A}$-reduced terms, we arrive at the DDs
\begin{align}
\begin{split}  \label{FDD}
F(\beta,\alpha;t) & = \pi \; \mathcal{A}  \; D(\beta,\alpha;t)^3  \Big[ (1-  \beta) m^2 \quad \quad \quad \quad 
\\ & - \beta \alpha^2 M_\pi^2 + (1-\beta) [(1-\beta)^2 - \alpha^2] t/ 4 \Big]
\end{split}\\ 
\begin{split}
G(\beta,\alpha;t) & = - \pi \; \mathcal{A} \; \alpha  \; D(\beta,\alpha;t)^3 \Big[ m^2  
\\ - M^2_\pi  (1 - & \beta - \alpha^2)  + [(1-\beta)^2 - \alpha^2] t / 4  \Big] \label{GDD},
\end{split}
\end{align} 
The contribution from $\mathfrak{A}$-reduced terms has the form of a $D$-term. Using Feynman parameters for the denominator
$\mathfrak{B}^2 \; \mathfrak{C}^2$ and suitable changes of variables, we arrive at the contribution to Eq.~\eqref{nspinor}
\begin{equation}
- \Delta^{\big[ \mu} \int_{-1}^{1} d\alpha  \frac{\pi \mathcal{A} \alpha^{n+1} (1 - \alpha^2)}{[m^2 - (1-\alpha^2) t/4]^2} 
 \Big(-\frac{\Delta}{2}\Big)^{\mu_{1}} \cdots \Big(-\frac{\Delta}{2}\Big)^{\mu_n \big]}
\end{equation} 
from which we can identify the $D$-term
\begin{equation}
D(\alpha;t) = \pi \mathcal{A} \frac{\alpha (1-\alpha^2)}{[m^2 - (1-\alpha^2) t/4]^2}. 
\end{equation}
Although strictly speaking, the $D$-term is a contribution to the $G$-DD, we shall treat it separately for ease.

Switching now to asymmetric variables, we have
\begin{align}
\begin{split}
F(x,y;t)  & =  \pi \; \mathcal{A} \;  D(x,y;t)^3  \Big[ (1  -x) m^2 
\\  - x (x & +  2 y - 1)^2 M_\pi^2 + (1-x) y(1-x-y)t \Big] \label{Fdd} 
\end{split}\\
\begin{split}
G(x,y;t)  & =  - \pi  \; \mathcal{A} \; (x + 2 y - 1) \; D(x,y;t)^3 \Big[ m^2 
\\ + y(1& -x-y) t  - M^2_\pi (1 - x - (x + 2y - 1)^2)  \Big] 
\label{Gdd} 
\end{split}\\
D(y;t) & = \pi \mathcal{A} y(1-y) (2 y - 1) \; [m^2 - y (1-y) t]^{-2} \label{Dy}  
\end{align}
where the function $D(x,y;t)$ is given by Eq.~\eqref{dxy}. Accordingly $F(x,y;t)$ is M\"unchen symmetric and $G(x,y;t)$ is 
antisymmetric, while $D(y;t)$ is antisymmetric about $y = 1/2$. Notice $F(x,y;t)$ in Eq.~\eqref{Fdd} is not that of 
Eq.~\eqref{DD}. Given the ambiguity inherent in defining $F$ and $G$ DDs (cf. Eq.~\eqref{Gsum}), there is no reason to believe 
the $F$s would be the same. In principle, we could construct a gauge transformation \cite{Teryaev:2001qm} to render the $F$s 
the same. This would enable identification of the missing $G$ function unique to section \ref{double}. We shall not pursue this 
tangential point.\footnote{Notice the contribution to the GPD from the 
$D$-term resembles that of $H_\text{inst}$ in Eq.~\eqref{Finst} but is not identical. Both terms originate from a reduction of the 
spectator's propagator. In the case of the $D$-term, the spectator's propagator is completely removed by the $\mathfrak{A}$-reduction. 
For $H_{\text{inst}}$, there is residual $x$-dependence stemming from the light-cone instantaneous propagator $\gp / 2 P^+ (1 - x)$.}

The function $F(x,y;t)$ satisfies the reduction relations: it reduces to the quark distribution via Eq.~\eqref{B} and
integrates to the form factor \eqref{C}---the latter can only be checked numerically. Lastly then it remains to see whether the 
DD-based GPD lines up with true GPD calculated in section \ref{form}. To construct the GPD we use the form 
of Eq.~\eqref{correct} modified to handle the $D$-term in Eq.~\eqref{Dy} separately 
\begin{multline} \label{recorrect}
H(x,\zeta,t) = \int_0^1 dz \int_0^{1-z} dy \; \delta(x - z - \zeta y) 
\\ \times \Big[ F(x,y;t) + \frac{\zeta}{2 - \zeta} G(x,y;t) \Big] + \frac{\theta[\xpp(1 - \xpp)]} {2(2 - \zeta)} D(\xpp;t)
\end{multline}
\begin{figure}
\begin{center}
\epsfig{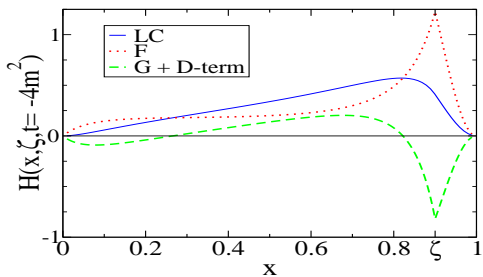}
\caption{Contributions to the covariant GPD from DDs. The light-cone GPD Eq.~\eqref{gpd} 
and the DD-based Eq.~\eqref{recorrect} are identical, denoted (LC) and plotted as a function of $x$ for fixed $\zeta = 0.9$ 
and $t = - 4 m^2$ for the mass $M_\pi = 0.15 m$. We also plot the individual contributions from $F$ in Eq.~\eqref{Fdd} and  
from $\frac{\zeta}{2 - \zeta} G + \frac{1}{2(2 - \zeta)} D$ in Eqs.~\ref{Gdd} and \ref{Dy}, denoted (G + D-term).}
\label{surprise3}
\end{center}
\end{figure}
In Figure \ref{surprise3}, we plot the GPD Eq.~\eqref{gpd} and the DD-based Eq.~\eqref{recorrect}. They are identical. 
We also plot the individual $F$ and $G + D$ contributions to the GPD. 
Even in the absence of the $D$-term, 
the contribution from $G$ cannot be neglected in ascertaining the DD. 

The contribution to the GPD from $F$ in the Figure is markedly different 
from Eq.~\eqref{gpd} and the analogous contribution from Eq.~\eqref{DD}.  These three GPDs, however, share the 
same form factor and quark distribution. Thus one can use DD-gauge freedom \cite{Teryaev:2001qm} to transform $F$ into a new function 
and throw away contributions from $G$. The result is an infinite set of different GPDs with identical form factors 
and quark distributions. As pointed out in section \ref{pc}, one must be careful to maintain positivity although
it is likely that there still is an infinite set of GPDs which would.

\section{Summary} \label{summy}
Above we consider two covariant models for pions: one with scalar constituents and the other with spin-$\frac{1}{2}$.
The spin-$\frac{1}{2}$ model requires regularization and we choose the method in \cite{Bakker:2000pk} in order to maintain positivity. 
For each case we derive the GPD from its matrix element definition which forces us to consider the triangle diagram for the
form factor with the plus momentum of the struck quark kept fixed in a general $\Delta^+ \neq  0$ frame.

We also construct the DDs for each model. The approach of \cite{Mukherjee:2002gb} leads only to one component 
of the DD (the ``forward-visible'' piece), thus resulting GPDs are incorrect and need not satisfy positivity.
This fact remains true even for $C$-odd distributions.  The ``gauge freedom'' inherent in defining $F$ vs. $G$ DDs
could be exploited, however, for phenomenological studies.  To obtain the other component of the DD, 
we calculate the matrix elements of twist-two operators. Resulting  DD-based GPDs then agree with those calculated 
on the light cone.

\begin{acknowledgments}
This work was funded by the U.~S.~Department of Energy, grant: DE-FG$03-97$ER$41014$.  
\end{acknowledgments}

\end{document}